\documentclass[twocolumn,showpacs,preprintnumbers,amsmath,amssymb]{revtex4} 
\input psfig.sty 
\def \be {\begin{equation}} 

\def \ee {\end{equation}} 
\def \bea {\begin{eqnarray}} 
\def \eea {\end{eqnarray}} 

\usepackage{graphicx}
\usepackage{dcolumn}
\usepackage{bm}
\usepackage{epsfig} 

\begin{document}
\title{Probing cosmic opacity at high redshifts with gamma-ray bursts}
\author{R. F. L. Holanda$^{1}$  } \email{holanda@uepb.edu.br}
\author{V. C. Busti$^{2}$} \email{vcbusti@astro.iag.usp.br}
\affiliation{$^1$ Departamento de F\'{\i}sica, Universidade Estadual da Para\'{\i}ba, 58429-500, Campina Grande - PB, Brasil,
\\ and Departamento de F\'{\i}sica, Universidade Federal de Campina Grande, 58429-900, Campina Grande - PB, Brasil
\\ $^2$ Astrophysics, Cosmology \& Gravity Centre (ACGC), and Department of Mathematics and Applied Mathematics, University of Cape Town, Rondebosch 7701, Cape Town, South Africa}

\begin{abstract}

\noindent Probing the evolution of the universe at high redshifts with standard candles is a powerful way to discriminate dark energy models, where  
an open question nowadays is whether this component is constant or evolves with time. One possible source of ambiguity in this kind of analyses comes from cosmic opacity, which can mimick a dark enery behaviour. However, most tests of cosmic opacity have 
been restricted to the redshift range  $z<2$.  In this work, by using luminosity distances of gamma-ray bursts (GRBs), { given the validity of the Amati relation}, and the latest $H(z)$ data we determine constraints on 
the cosmic opacity at high redshifts ($z>2$) for a flat $\Lambda$CDM model. A possible degenerescence  of the results with the adopted cosmological model is also investigated 
by considering a flat XCDM model. 
The limits on cosmic opacity in the redshift range $0<z<2$ are updated with type Ia supernovae (SNe Ia) from the Union2.1 sample, where we 
added the most distant ($z=1.713$) spectroscopically confirmed SNe Ia. From the analyses performed, we find that both SNe Ia and GRBs samples  are compatible with a 
transparent universe at $1\sigma$ level and the results are independent of the dark energy equation of state parameter $w$.
\end{abstract}

\maketitle

\section{Introduction}\label{sec:introduction}

Type Ia supernovae (SNe Ia) observations provide the most direct evidence for the current cosmic acceleration. In the context of Einstein's general theory of relativity, 
this result implies either the existence of some sort of dark energy, constant or that varies slowly with time and space (see \cite{caldwell,li} 
for recent reviews), or that the matter content of the universe is subject to dissipative processes \cite{Lima1999,chimento}. On the other hand, since SNe  Ia observations are affected by at least four different sources of opacity, namely, the Milky Way, the hosting galaxy,  intervening galaxies, and the intergalactic  medium, alternative mechanisms contributing to the acceleration evidence or even mimicking the dark energy behaviour have been proposed. Examples are possible evolutionary effects  in SNe Ia events \cite{drell,combes}, a local Hubble bubble \cite{zehavi,conley}, modified gravity \cite{Ishak,ks2007,Bertschinger},  unclustered sources of light 
attenuation \cite{Aguirre,Rowan,Goobar} and the existence of Axion-Like-Particles (ALPs), arising in a wide range of well-motivated  
high-energy physics scenarios, and that could lead to dimming of SNe Ia brightness (\cite{Av2}, for a review of the phenomenology  
of the weakly-interacting-sub-eV-particles (WISPs) see \cite{Jaeckel}).

In the last years, the  cosmic distance duality (CDD) relation has been used to verify the existence of exotic physics as well as the presence of opacity and systematic errors in SNe Ia 
observations. The CDD relation  connects the luminosity distance, $D_{\scriptstyle L}$, to the angular diameter distance, $D_{\scriptstyle A}$, by \cite{eth33,ellis07}
\begin{equation}
 \frac{D_{\scriptstyle L}}{D_{\scriptstyle A}}{(1+z)}^{-2}=1.
 \label{rec}
\end{equation}
This relation is valid for all cosmological models based on Riemannian geometry, requiring only that source and observer are connected by null geodesics in a Riemannian spacetime,  
that the number of photons is conserved \cite{ellis07} and narrow-beam effects are negligible \cite{busti2012,clarkson}. 

Several methods have been proposed in the literature to test modifications of the CDD relation. For instance, \cite{bk04} used SNe Ia data as measurements  
of $D_{\scriptstyle L}$  and estimates of $D_{\scriptstyle A}$ from FRIIb radio galaxies \cite{daly} and ultra-compact radio 
sources \cite{G94} to test possible deviations of the CDD relation. They found a 2$\sigma$ violation caused by an excess brightening of SNe Ia at $z > 0.5$, perhaps due to 
lensing  magnification bias. \cite{h3,h2,h3a} proposed a test of the CDD relation by using luminosity distances of SNe Ia  and angular diameter distances 
of galaxy clusters. The verification of the CDD relation validity depended on the assumptions used to describe the galaxy clusters and the light-curve fitters  used in SNe Ia 
analysis (see also \cite{Nair,xi,nam}). 

Still with SNe Ia and angular diameter distances of galaxy clusters data,  \cite{li2013} explored the impact of cosmic opacity on the cosmic acceleration observation. 
Their results suggested that an  accelerated cosmic expansion is still needed to account for the dimming of SNe Ia and the standard cosmological scenario remains to be supported 
by current observations. By using four different parameterizations to possible CDD deformations and SNe Ia data \cite{Amanullah} the authors of ref. \cite{Lima} found that only if the cosmic opacity is fully negligible,  the description of an accelerating universe powered by dark energy or some alternative gravity theory must be invoked. { In Ref. \cite{HAC} measurements of the expansion rate $H(z)$ and Union 2.1 SNe Ia compilation were used to impose cosmological model-independent constraints on cosmic opacity \cite{foot}. It was found that a completely transparent universe  is in agreement with the largest SNe Ia sample.} 

More recently, an interesting test of the CDD relation via cosmic microwave background spectrum was proposed by \cite{ellis2013}. As a result, 
it was found that the CDD  
relation cannot  be violated by more than 0.01\% between the decoupling era and today.  Furthemore, a cosmological model-independent test involving only measurements of the 
gas mass fraction of galaxy clusters from Sunyaev-Zeldovich and X-ray surface brightness observations was discussed by \cite{h5b}, where  
no significant  violation of the CDD relation was found, supporting a transparent universe in the X-ray and radio wavelengths.

In face of the discussed above, most tests of cosmic opacity  have been limited in the redshift range $0 <  z < 1.7$. On the other hand, since current SNe Ia data even when 
combined with all other currently available data sets cannot yet determine whether the energy density of dark energy is constant as required by the $\Lambda$CDM model or  
time-varying, some authors have used luminosity distances of gamma-ray bursts (GRBs) to explore the universe at higher redshifts ($2<z<8$) to understand how dark energy behaves. 
GRBs provide a means to test cosmological models once phenomenological relations are used to transform them into stardardizable candles 
(see \cite{amati,ghirlanda}). For example, they were used to test dark energy models \cite{schaefer2007,wang,ratra2010,wei2} 
and also possible bias from underdense lines of sight \cite{busti_prd_2012}. In this way, in order to have confidence in analyses involving GRBs it is important to verify the  
existence of cosmic opacity also in high redshifts.

In this paper, we  probe cosmic opacity at higher redshifts through luminosity distances from GRBs data, { by assuming the validity of the Amati relation \cite{amati}}, and with the latest 26 opacity-free  measurements of the Hubble expansion 
in the  range ($0.1<z<2.30$) using the approach proposed by Avgoustidis {\it et al.} \cite{Av,Av2} (see next section to details). We find that GRB data are in a very good agreement 
with a transparent universe, with bounds consistent with the ones derived with SNe Ia data. A possible degenerescence of the opacity constraints with the underlying cosmological 
model is also investigated, where it is shown that the opacity parameter $\epsilon$ is nearly insensitive of a different dark energy equation of state parameter $w$ for a flat XCDM model.

The paper is organized as follows. In Sec. II we describe the methodology. In Sec. III  the observational data we use in the statistical analyses are presented. The corresponding
constraints on the cosmological opacity  are investigated in Sec. IV. The article is finished with a summary of the the main results in the conclusion section.

\section{Luminosity distance and cosmic opacity}

\subsection{Methodology}

\begin{figure*}[t]
\label{Fig1}
\centerline{
\psfig{figure=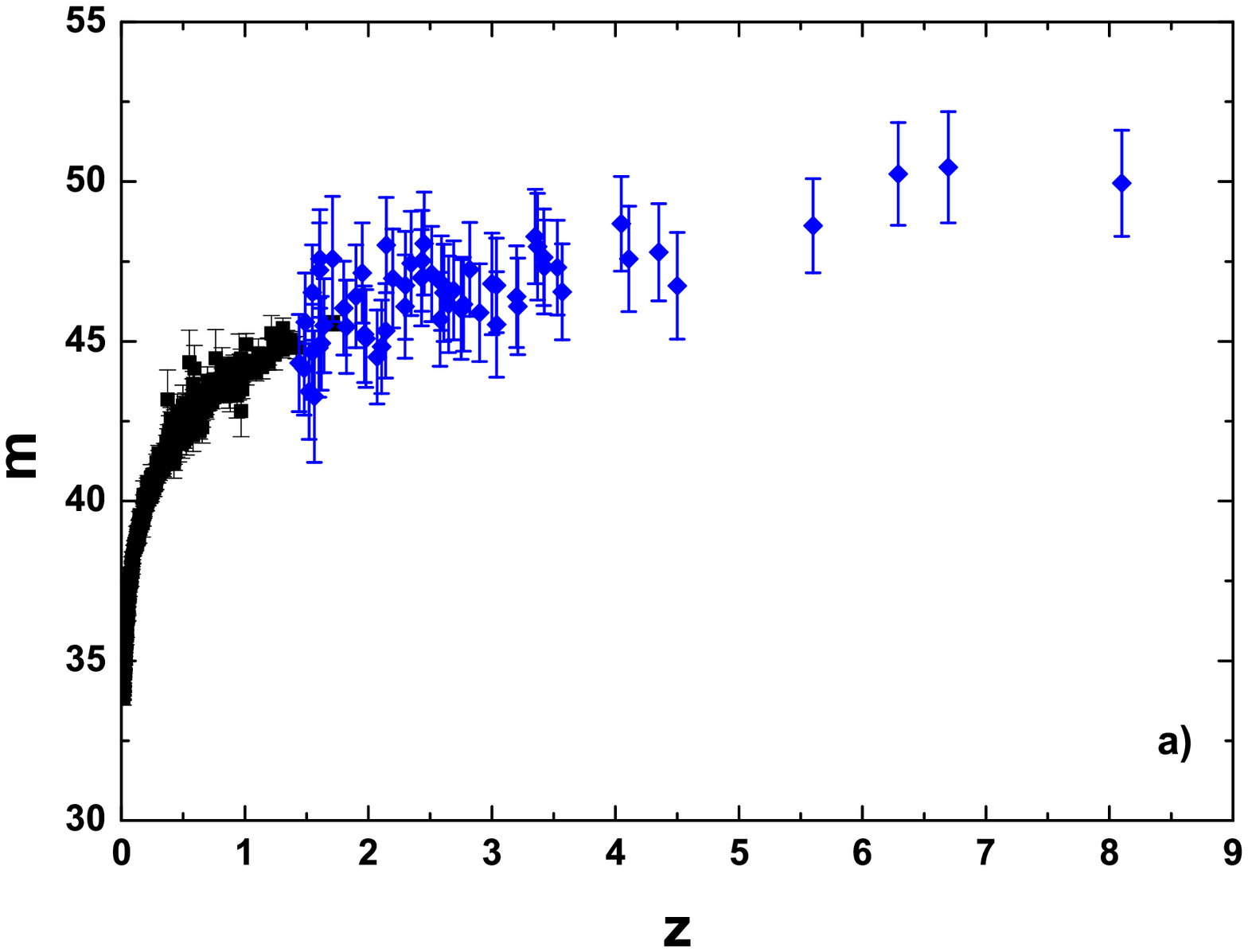,width=0.5\textwidth}
\psfig{figure=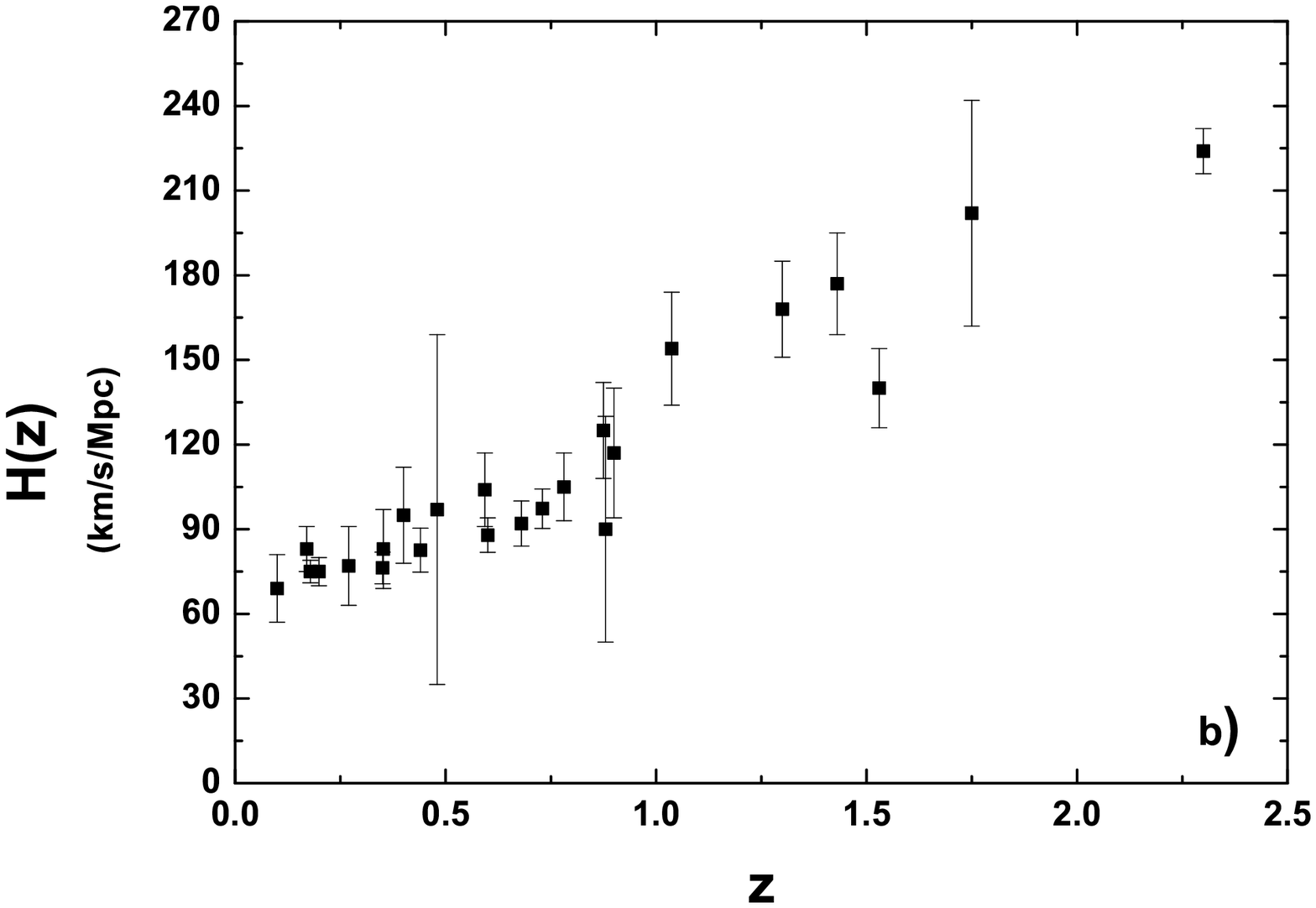,width=0.5\textwidth}
\hskip 0.1in}
\caption{{{a)}} Distance modulus $m$ as a function of redshift for 580 SNe Ia from the Union2.1 sample \cite{suzuki} and one high-redshift SNe Ia \cite{Rubin} 
(black squares) plus 59 GRBs (blue diamonds) 
calibrated by Wei (2010). {{b)}} 19 $H(z)$ measurements from cosmic chonometers \cite{simon,stern,moresco} plus 7 $H(z)$ measurements from 
baryon acoustic oscillations (BAOs) \cite{blake,reid,xu,busca,chuang}.}
\end{figure*}

The methodology used in our analysis was proposed by \cite{Av}. As pointed by these authors, the distance modulus derived from SNe Ia or, in our case, 
from GRBs, is  systematically affected if there were a source of ``photon  absorption" affecting the universe transparency. Any effect that reduces the number of photons would 
dim the brightness of the source and increases its $D_L$. Thus, if $\tau(z)$ denotes the opacity  between an observer at $z=0$ and a source at $z$ due to, e.g., extinction, the 
flux received from the source would be attenuated by a factor $e^{-\tau(z)}$ and thus the observed luminosity distance ($D_{L, obs}$) is related with the true luminosity 
distance ($D_{L, true}$) by
 \begin{equation}
 D_{L, obs}^2=D_{L,true}^2 e^{\tau(z)} \, .
 \end{equation}
Therefore, the \emph{observed} distance modulus is given by \cite{chen2009a,chen2009b}
\begin{equation}
\label{distancemod}
m_{obs}(z)=m_{true}(z)+2.5(\log e) \tau(z) \, .
\end{equation}
 
As it is largely known, for a flat Friedmann - Lema\^{i}tre - Robertson - Walker (FLRW) cosmology \cite{Hogg}
\begin{equation}
 d_{L,true}(z)=(1+z)c\int_o^z\frac{dz'}{H(z)},
 \end{equation}
 where $c$ is the speed of light and
  \begin{equation*}
  H(z)=H_0 E(z, \Omega_M,w)\\, 
  \end{equation*}
  \begin{equation}
  E(z,\Omega_M,w)=[\Omega_M(1+z)^3+(1-\Omega_M)(1+z)^{3(1+w)}]^{1/2}.
  \end{equation}
In the above expressions, $\Omega_M$ stands for the matter density parameter measured today and $w$ for the dark energy equation of state parameter.  
If $w=-1$ we have the so-called  flat $\Lambda$CDM model.  Avgoustidis {\it et al.} \cite{Av,Av2}, to be able to use the full redshift range of the available data, considered the following  parameterization of a deviation from the Etherington relation  $D_L=D_A(1+z)^{(2+\epsilon)}$, with $\epsilon$ parameterizing departures from transparency. To understand the physical meaning of a constraint on $\epsilon$  these authors argued that for small $\epsilon$ and $z \leq 1$ this is equivalent to assuming an optical depth parameterization  $\tau(z)=2\epsilon z$ or  $\tau=(1+z)^{\alpha}-1$ with the correspondence $\alpha=2 \epsilon$. 
  
In our analysis, in order to obtain tighter limits on cosmic opacity, we consider a simple linear parameterization for $\tau(z)=\epsilon z$ and measurements of $m_{obs}$ (or $D_{L, obs}$) are taken from the GRB compilation  (see Sec. III) to probe the cosmic opacity  on a redshift range which has not been explored yet \cite{footnote}. The unknown parameters $\Omega_M$, 
$\epsilon$ and $w$ are obtained by fitting the GRB data  separately and jointly  with $H(z)$ measurements on two flat cosmic scenarios, namely,  
$\Lambda$CDM and XCDM. The basic  
idea behind is that while the brightness of the GRB can be affected by at least four different sources of opacity (the Milky Way, the hosting galaxy,  intervening galaxies, and the  
intergalactic medium)  the $H(z)$ measurements  are obtained either from ages of old passively evolving galaxies and rely only on the detailed shape of the galaxy spectra but not on the  
galaxy luminosity or from the features in baryonic acoustic oscillation which are completely independent from the measured flux. Therefore, $H(z)$ values are not affected by a 
non-zero  $\tau(z)$ since $\tau$ is assumed  not to be  strongly wavelength dependent in the optical band.  
In order to compare and update previous results, we also consider SNe Ia observations from the Union2.1  sample \cite{suzuki}, where we add  the most distant ($z=1.713$) spectroscopically confirmed SNe Ia \cite{Rubin}.

\section{Data sets}

\subsection{Gamma-ray bursts}

GRBs are the most violent explosions of the Universe. Due to their brightness, they are measured up to very far distances, where the most distant GRB was detected at $z=8.2$ \cite{highzgrb} {(a photometric determination of the redshift of a GRB at $z=9.4$ was also claimed \cite{highzgrb2})}.
This is far beyond what is covered by SNe Ia, which reach $z \sim 1.7$. Therefore, the possibility to use GRBs as standardized candles may allow cosmology tests to be performed 
in a redshift region not probed by any other sources.

In order to do so, many phenomenological relations to calibrate GRBs for cosmological purposes appeared in the literature, where one of the most successful proposals is known as the 
Amati relation \cite{amati}. It relates the isotropic-equivalent radiated energy in gamma-rays $(E_{iso})$ and the photon energy at which the $\nu F_{\nu}$ is brightest $(E_{peak})$, 
$\nu$ standing for the frequency and $F_{\nu}$ for the flux at frequency $\nu$. The Amati relation can be expressed by a power law: $E_{p,i} = a\times E_{iso}^b$, where 
$E_{p,i}= E_{peak}(1+z)$ is the cosmological rest-frame peak energy and $E_{iso}$ is given by

\begin{equation}
 E_{iso} = 4\pi D_L^2 S_{bolo} (1+z)^{-1},
\end{equation}
$S_{bolo}$ denoting the bolometric fluence in gamma-rays in a given GRB.

Unfortunately, the GRBs detected at very low redshifts seem to belong to a different class of objects when compared to their high-redshift counterparts, which makes the  calibration 
process tricky. One way around this problem was proposed by \cite{kodama} and \cite{liang}, which consists of using SNe Ia to calibrate the GRBs. 
{ The method was updated by 
Wei \cite{wei}, who used the 557 SNe Ia from the Union2 sample \cite{Amanullah} and 109 GRBs. For each GRB which is in the same redshift range of the SNe Ia, $0 \leq z \leq 1.4$, were found the four closest SNe Ia and a cubic interpolation was applied to derive its distance modulus. 
Once the 50 GRBs in this redshift range had their distance modulus determined, a fit was performed to obtain the parameters $a$ and $b$ in the Amati relation, which in its turn allowed one to derive the distances for the remaining high-redshift GRBs. After this process, we are left with 59 GRBs with $z > 1.4$ which can be used for cosmological purposes (shown in Fig. 1a as blue diamonds)}. Note that the GRBs are calibrated with the SNe Ia without a correction for the opacity in SNe Ia data. Therefore, if a non-zero opacity is present its effect should become even stronger when analyzing the GRB data.  

At this point, it is important to remember that GRBs do not have the same status of other cosmological probes (e.g. SNe Ia, baryon acoustic oscillations, cosmic microwave background
anisotropies). This is due to our poor knowledge of the physical processes driving the explosion. For example, there is a strong debate in the literature concerning whether the Amati relation is an intrinsic property of GRBs or merely a combination of selection effects (see \cite{atteia}, and references therein). { Moreover, \cite{salvaterra} detected a strong evolution in the luminosity function of GRBs. This may indicate that GRBs are intrinsically more luminous at high redshifts. Since we used the low-redshift GRBs to calibrate the high-redshift ones, we would infer a smaller distance to the GRB, so a smaller value for the opacity. Therefore, if this effect is important we are underestimating the opacity.}

\subsection{H(z) measurements}

In recent years $H(z)$ measurements of the Hubble parameter have been used to constrain several cosmological parameters \cite{simon,stern,chen2011,seik,moresco,Farooq}. 
In this paper, we use 19 $H(z)$ measurements from cosmic chonometers \cite{simon,stern,moresco} plus 7 $H(z)$ measurements from 
baryon acoustic oscillations (BAOs) \cite{blake,reid,xu,busca,chuang} 
in the redshift range $ 0 < z < 2.3$. 
The $H_0$ influence on our results is explored by considering a prior in the analysis: $68 \pm 2.8$ km s$^{-1}$ Mpc$^{-1}$, from a median statistics analysis  of 553 measurements of $H_0$ \cite{chen2011}, 
consistent with the value derived recently
by {\it Planck} (Planck Collaboration 2013) \cite{Planck}. Moreover, following \cite{Farooq}, 
the distribution of $H_0$ is assumed to be Gaussian with one  standard deviation width  $\sigma_{H_0}$ and mean $\bar{H_0}$, in this way, one can  build the posterior likelihood 
function $\mathcal{L}_{H}(\textbf{p})$ that depends only on the parameters $\textbf{p}$ by integrating the product of exp$(-\chi_H^2 /2)$ and the $H_0$ prior likelihood function
exp$[-(H_0-\bar H_0)^2/(2\sigma^2_{H_0})]$,
\begin{equation}
\label{eq:likely}
\mathcal{L}_{H}(\textbf{p})=\frac{1}{\sqrt{2\pi \sigma^2_{H_0}}}
   \int \limits_0^\infty e^{-\chi_H^2(H_0,\textbf{p})/2}
   e^{-(H_0-\bar H_0)^2/(2\sigma^2_{H_0})} dH_0.
\end{equation}
The results from $H(z)$ measurements are obtained by maximizing the likelihood $\mathcal{L}_H(\textbf{p})$,
or equivalently minimizing $\chi_H^2(\textbf{p}) = -2
\mathrm{ln}\mathcal{L}_{H}(\textbf{p})$, with respect to the
parameters $\textbf{p}$ to find the best-fit parameter values for flat $\Lambda$CDM ($\textbf{p}=\Omega_M$) and XCDM ($\textbf{p}=\Omega_M, w$) universes.

\subsection{SNe Ia Sample}

Here we consider the 580 SNe Ia compiled by Suzuki et al. (2012) \cite{suzuki}, known as the Union2.1  sample. The sample is in the redshift range $0.015<z<1.43$, where we add 
the most distant ($z=1.713$) spectroscopically confirmed SNe Ia \cite{Rubin}. This SNe Ia sample was calibrated with SALT2 \cite{guy} light-curve fitter and is plotted 
in Fig. 1(a) (black squares). 

\begin{figure*}[t]
\label{Fig2}
\centerline{
\psfig{figure=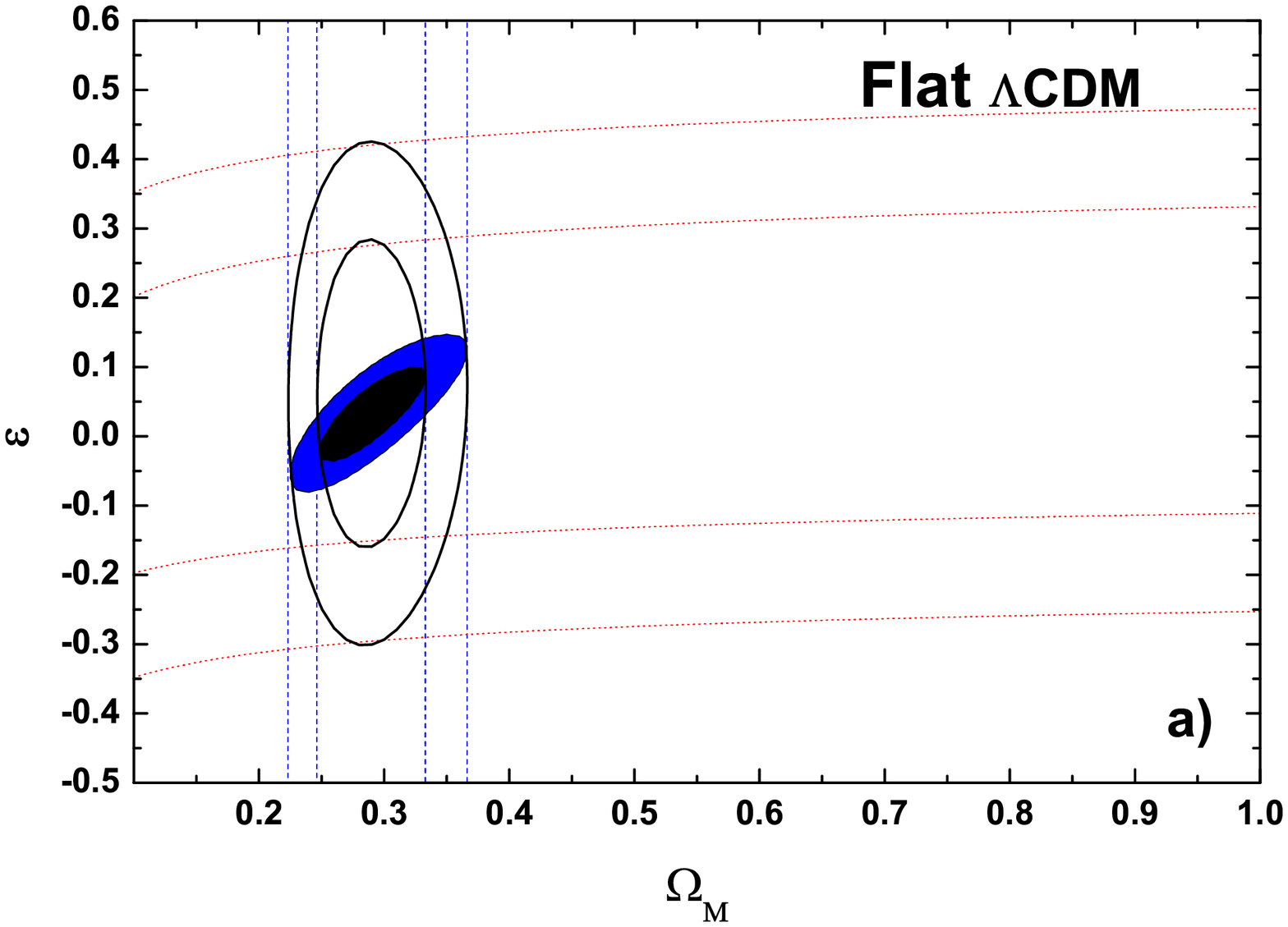,width=0.5\textwidth}
\psfig{figure=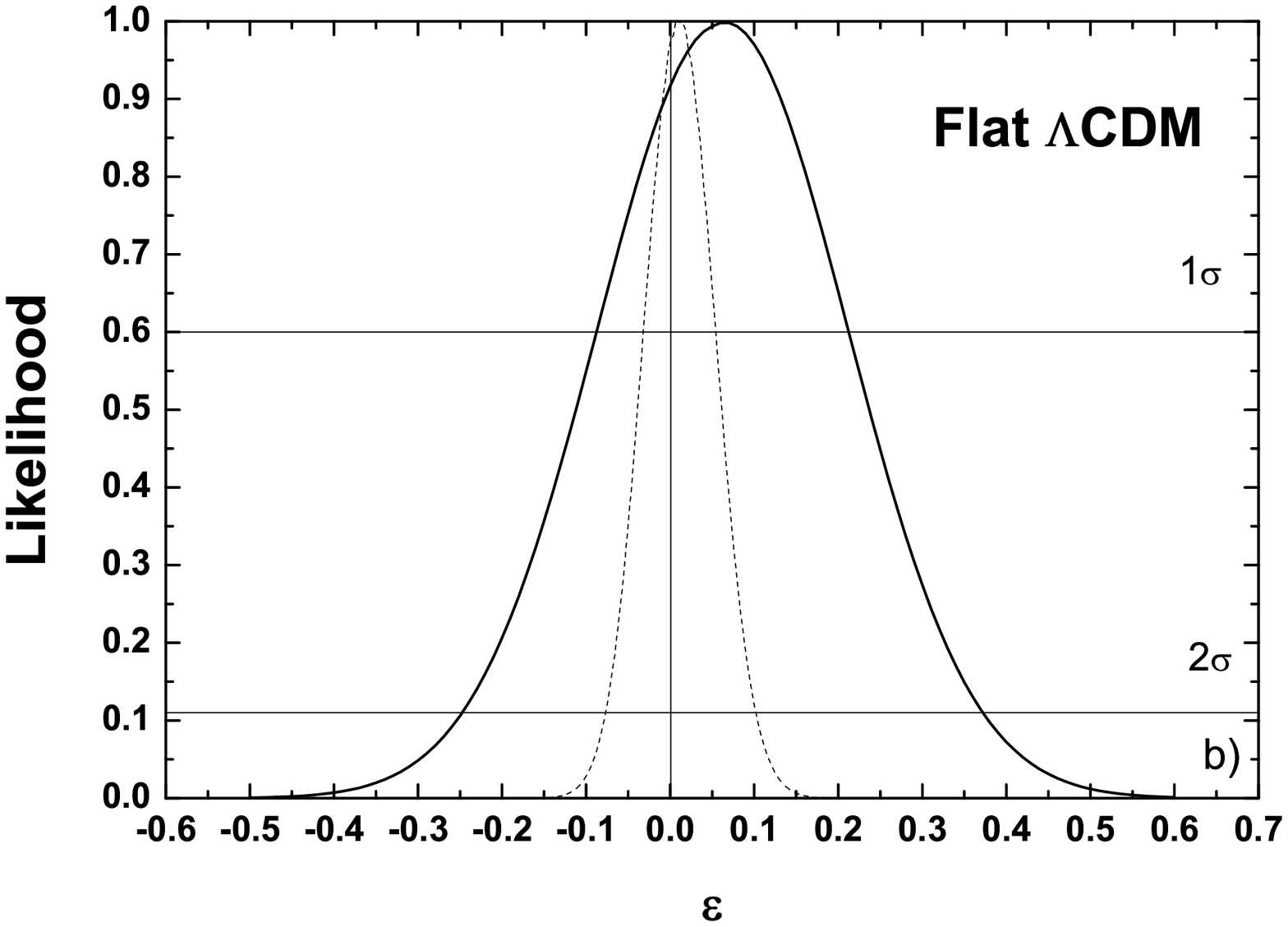,width=0.5\textwidth}
\hskip 0.1in}
\caption{a) Contours in the ($\epsilon, \Omega_M$) plane for 1 and 2$\sigma$ confidence levels. The dashed blue and dotted red curves correspond to, respectively, limits by 
using $H(z)$ and GRB data  separately. 
The black curves corresponds to limits of the $H(z)$ + GRB analysis. The inner filled contours show the constraints obtained by using SNe Ia + $H(z)$ data. 
b) Likelihood of the $\epsilon$ parameter. The solid and dotted lines correspond to analysis of the $H(z)$ + GRB and $H(z)$ + SNe Ia, respectively.}
\end{figure*}

\subsection{Analysis and results}

We obtain the constraints to the set of parameters ${\mathbf{p}}$ by evaluating the likelihood distribution function, ${\cal{L}} \propto e^{-\chi^{2}/2}$, with

\begin{equation}
\chi^{2} =  \sum_{z}\frac{[m_{obs}(z) - m_{true}(z,\mathbf{p} )-1.08\epsilon z]^2}
{\sigma^2_{m(obs)}} + \chi^2_{H}(z,\mathbf{p})
\end{equation}
where  $\sigma^2_{m(obs)}$ is the error associated to distance modulus from GRB (or SNe Ia) and $\chi^2_{H}(z,\mathbf{p})$ is given by Eq. (7). $m_{true}$ is obtained 
via $m_{true}=5\log_{10} D_{L,true} + 25$, while $D_{L,true}$ is given by equation (4) with $w=-1$ for a flat $\Lambda$CDM and $w$ as a free parameter for a flat XCDM model. As it is largely employed in the literature, all the results in our analysis  from SNe Ia and GRB data are derived by marginalizing the likelihood function over the pertinent nuisance parameters \cite{riess}.

\subsubsection{Flat $\Lambda$CDM}

In Fig. 2(a) we  show contours of $\Delta \chi^2 = 2.30$ $(1\sigma)$ and 6.17 $(2\sigma)$ on the $\Omega_M -\epsilon$ plane when the GRBs, SNe Ia and $H(z)$ samples are considered.
For the GRB sample (red dotted curves) we find that a perfect transparent universe ($\epsilon=0$) is allowed by the current data with $\epsilon=0.09\pm 0.25$ ($1\sigma$), 
however, $\Omega_M$ is not limited. This means that using only the GRB sample we cannot constrain simultaneously the energy content of
the flat $\Lambda$CDM model and the $\epsilon$ parameter. On the other hand, $H(z)$ data do not constrain  $\epsilon$ (dashed blue curves), but impose restrictive 
limits to $\Omega_M$, such as $\Omega_M=0.28 \pm 0.04$ $(1\sigma)$. Thus, more stringent constraints on the parameter space ($\Omega_M -\epsilon$) can be obtained by combining GRB + $H(z)$ data. 
The black contours show the 1 and 2$\sigma$ bounds on the $\Omega_m -\epsilon$ plane from the  GRB + $H(z)$ joint analysis, which provides $\epsilon=0.06\pm 0.20$ 
and $\Omega_M=0.28 \pm 0.04$ at $1\sigma$ confidence level. Further, we update the constraints on $\epsilon$ using
the latest SNe Ia compilation \cite{suzuki} plus the most distant ($z=1.713$) spectroscopically confirmed SNe Ia \cite{Rubin}. By combining SNe Ia $+$ $H(z)$ data we 
have $\epsilon=0.02\pm 0.06$ and $\Omega_M=0.28 \pm 0.04$ $(1\sigma)$ (filled contours).   

By marginalizing over $\Omega_M$, panel 2(b) displays the likelihood for the $\epsilon$ parameter with GRB $+$ $H(z)$ and SNe Ia $+$ $H(z)$ analyses. We 
obtain $\epsilon=0.06\pm 0.18$ and $\epsilon=0.020 \pm 0.055$, respectively at $1\sigma$ level. As one may conclude, our results support a transparent universe.

\subsubsection{Flat XCDM}

\begin{figure*}[t]
\label{Fig3}
\centerline{
\psfig{figure=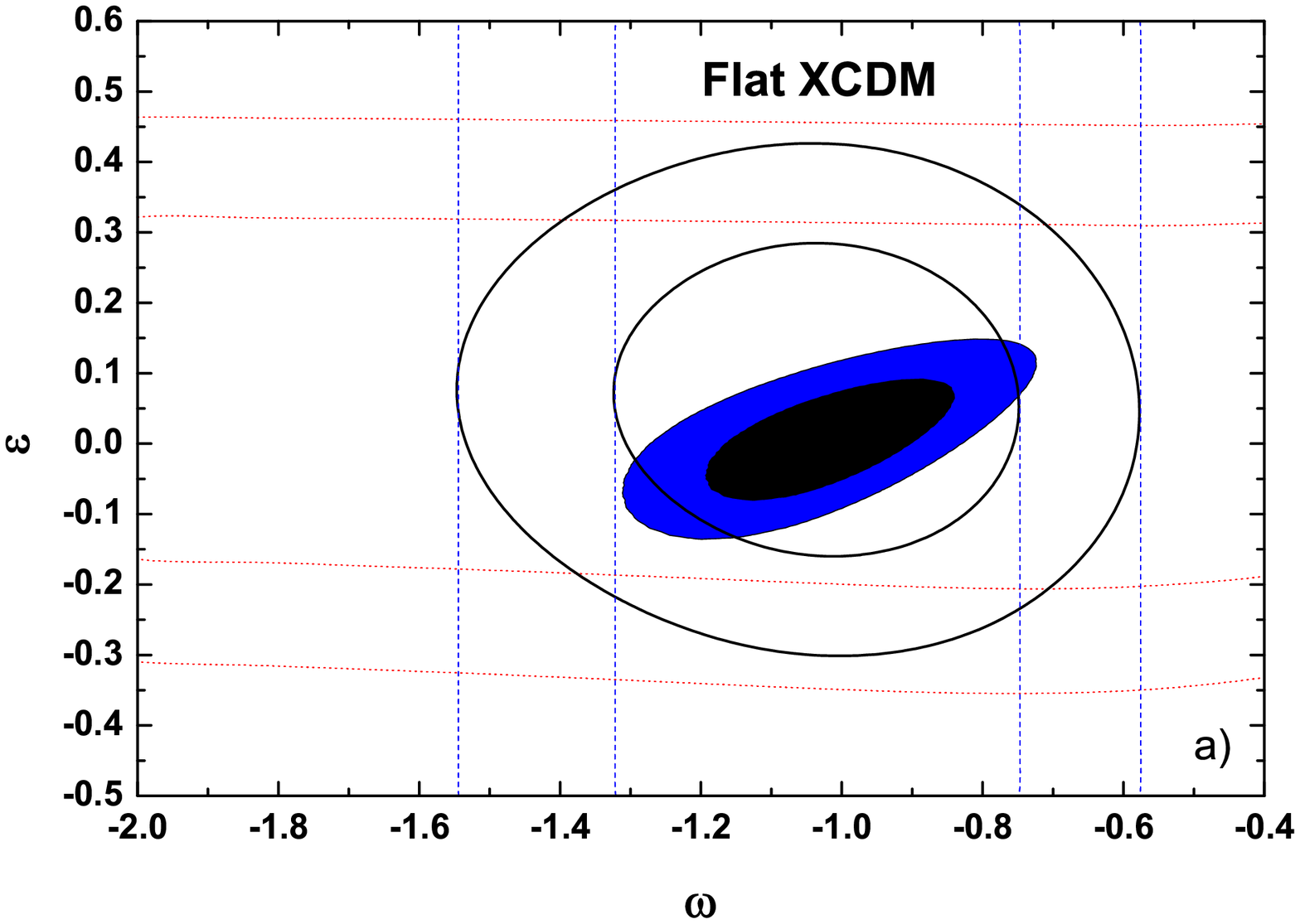,width=0.5\textwidth}
\psfig{figure=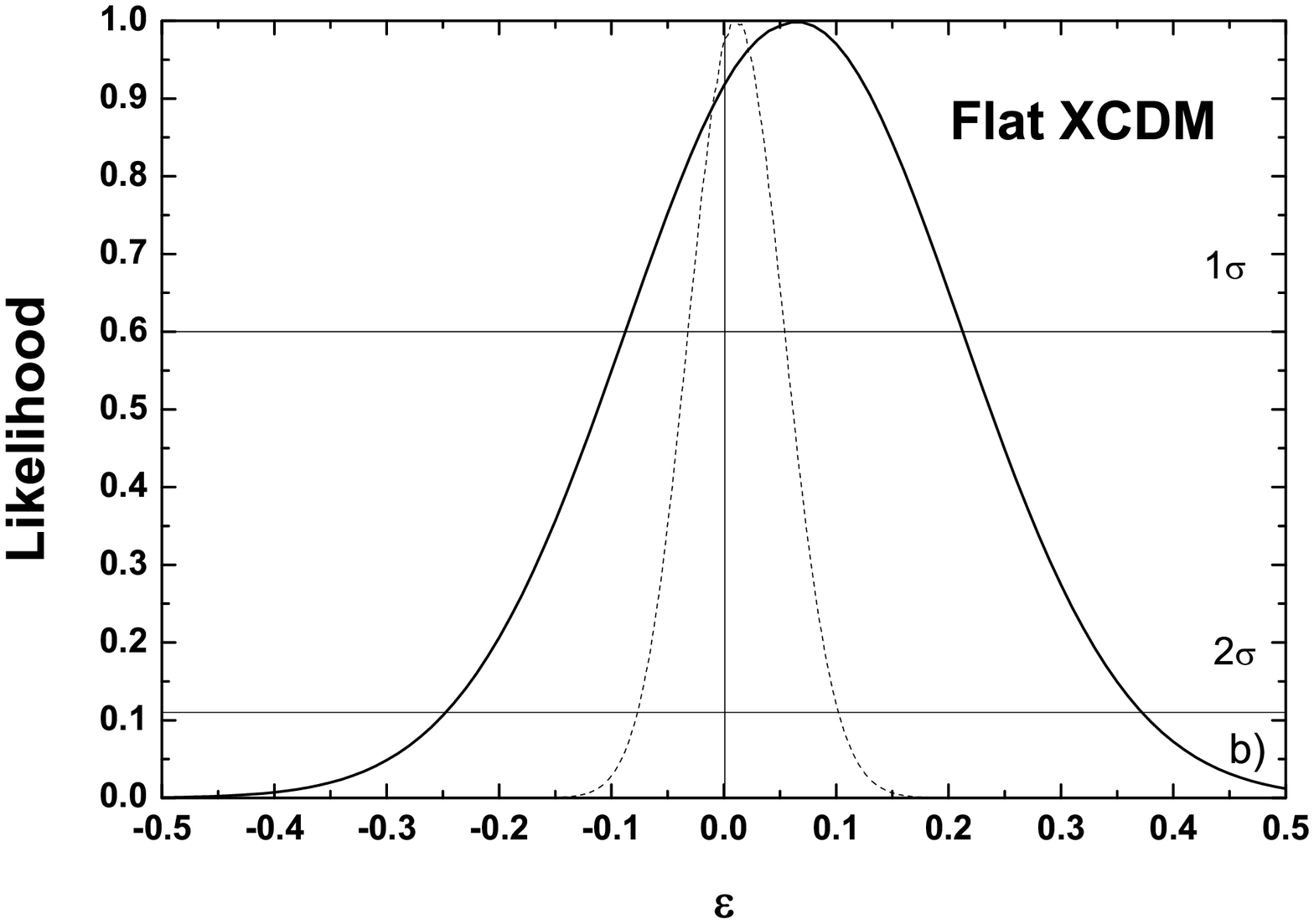,width=0.5\textwidth}
\hskip 0.1in}
\caption{a) Contours in the ($\epsilon, w$) plane for 1 and 2$\sigma$ confidence levels. The dashed blue and dotted red curves correspond to, respectively, limits by using $H(z)$ data and GRB separately. Note that 
the constraints from GRB on $\epsilon$ are independent of the $w$. The black curves corresponds to limits of the $H(z)$ + GRB analysis. The inner filled contours show the constraints 
obtained by using SNe Ia + $H(z)$ data.  b) Likelihood of the $\epsilon$ parameter. The solid and dotted lines correspond to analysis of the $H(z)$ + GRB and $H(z)$ + SNe Ia, respectively.} 
\end{figure*}

In Fig. 3(a) we  show 1 and 2$\sigma$ contours on the $w -\epsilon$ plane, marginalizing over $\Omega_M$ and considering the GRB, SNe Ia and $H(z)$ samples.
It is important to note that the H(z) measurements derived from BAOs were obtained within the $\Lambda$CDM model, and therefore are model dependent. Nonetheless,  we verified they do not change the  constraints for $\epsilon$, providing stronger constraints to $w$ since the data were generated with $w=-1$.
As a general result, the limits on $\epsilon$ in the ($\epsilon$, $w$) plane are wider than in the flat $\Lambda$CDM model, but the value of $\epsilon$ is independent of $w$. 
For the GRB sample (red dotted curves), we find that a perfect transparent universe ($\epsilon=0$)  is 
allowed by the current data with $\epsilon=0.10 \pm 0.22$ ($1\sigma$) and the 1$\sigma$ and 2$\sigma$ contours are independent of $w$, however, $w$ is not limited. 
The dashed blue curves show constraints from the $H(z)$ sample,  which imposes  limits only on $w$, such as $w=-1.10 \pm 0.35$ $(1\sigma)$. In this way, better constraints on the 
parameter space ($w -\epsilon$) are obtained by combining GRB + $H(z)$ data. The black contours show 1 and 2$\sigma$ confidence levels on the $w -\epsilon$ plane from the 
GRB $+$ $H(z)$ joint analysis, which provides $\epsilon=0.06\pm 0.23$ and $w=-1.1 \pm 0.35$ $(1\sigma)$. 

At this point, it is very important to comment that previous works \cite{Av,Av2} did not explore the ($\epsilon$, $\omega$) plane using SNe Ia.
In Fig 3(b) we show  the constraints on $\epsilon$ and $w$ using the latest SNe Ia compilation \cite{suzuki} plus the most distant ($z=1.713$) spectroscopically confirmed 
SNe Ia \cite{Rubin}, jointly with $H(z)$ data. We obtain $\epsilon=0.015\pm 0.090$ and $w=-1.01 \pm 0.18$ $(1\sigma)$ (filled contours). By marginalizing over $w$ and $\Omega_M$, 
panel 3(b) displays the likelihood for $\epsilon$ with GRB $+$ $H(z)$ and SNe Ia $+$ $H(z)$ analyses. We obtain $\epsilon=0.057\pm 0.21$ and $\epsilon=0.015\pm 0.060$, 
respectively, at $1\sigma$ level. This result is in full agreement with those of the flat $\Lambda$CDM analysis. { In table I  we show some recent constraints on the cosmic opacity by using approaches involving SNe Ia and $H(z)$ observations as well as our results.}

\begin{table*}
\caption{Constraints on $\epsilon$. The values obtained of the analyses with SNe Ia + $H(z)$ are from the range $0<z<2$. On the other hand, the values obtained  of the analyses with GRB + $H(z)$ are from the range $1.4<z<8.1$.}
{\begin{tabular} {c||c||c|c|c}
Reference & Data set & Model & $\tau(z)$&$\epsilon$ ($1\sigma$) 
 \\
\hline \hline 
Avgoustidis et al. 2009& 307 SNe Ia + 10 $H(z)$ & flat $\Lambda$CDM &  $\tau(z)=2\epsilon z$ &$-0.01^{+0.06}_{-0.04}$ \\
Avgoustidis et al. 2010& 307 SNe Ia + 12 $H(z)$&flat $\Lambda$CDM&  $\tau(z)=2\epsilon z$ &$-0.04^{+0.04}_{-0.03}$  \\
Holanda et al. 2013 &    581 SNe Ia + 28 $H(z)$ & model independent&  $\tau(z)=2\epsilon z$&$0.017^{+0.052}_{-0.052}$ \\
This paper& 581 SNe Ia + 19 $H(z)$ & flat $\Lambda$CDM &  $\tau(z)=\epsilon z$&$0.02^{+0.055}_{-0.055}$\\
This paper& 59 GRB +   19 $H(z)$ & flat $\Lambda$CDM & $\tau(z)=\epsilon z$ &$0.06^{+0.18}_{-0.18}$\\
This paper& 581 SNe Ia +19 $H(z)$ & flat XCDM & $\tau(z)=\epsilon z$ &$0.015^{+0.060}_{-0.060}$\\
This paper& 59 GRB + 19 $H(z)$ &flat XCDM &  $\tau(z)=\epsilon z$&$0.057^{+0.21}_{-0.21}$ \\

\hline
\end{tabular}} \label{ta2}
\end{table*}

\section{Conclusions}

In the last few years, several cosmological observables, such as baryon accoustic oscillations, luminosity distances of SNe Ia, angular diameter distances of galaxy clusters, 
gas mass fraction and $H(z)$ measurements have been used to probe cosmic opacity and/or searching for some evidence of new physics. However, due to limitations in the 
redshift distribution of the data, most of these analyses were limited in the redshift range $0<z<2$, where the region between the SNe Ia and the cosmic microwave background data
remained unexplored.   

In this work, we have probed cosmic opacity in the the redshift range $1.5<z<8$ through luminosity distances from GRBs, { by assuming the validity of the Amati relation \cite{amati}}, and the latest 26 measurements of the Hubble 
expansion from passively evolving galaxies and baryon acoustic oscillations in the range ($0.1<z<2.30$). In order to test a dependence of the method
with the adopted cosmology, we  compared the constraints from flat $\Lambda$CDM model with the flat XCDM model. We found that GRB data are in full agreement with a perfect transparent 
universe and the results are independent of $w$. We parameterized the cosmic opacity such as $\tau(z)=\epsilon z$. By marginalizing over $\Omega_M$, we 
obtained $\epsilon=0.06\pm 0.18$ $(1\sigma)$ (flat $\Lambda$CDM) and, by marginalizing over $\Omega_M$ and $w$, we obtained $\epsilon=0.057\pm 0.21$ at $1\sigma$ level for a flat 
XCDM model. We also 
used the Union2.1 SNe Ia sample, where we added the most distant ($z=1.713$) spectroscopically confirmed SNe Ia to impose limits on opacity in redshift range $0<z<2$. From 
the joint analyses involving SNe Ia and $H(z)$ data we got $\epsilon=0.020 \pm 0.055$ $(1\sigma)$ (flat $\Lambda$CDM) and $\epsilon=0.015\pm 0.060$ $(1\sigma)$ (flat XCDM). We believe 
that the results presented here reinforce the interest in the observational search for GRBs. When larger samples with smaller statistical and systematic 
uncertainties become available one may improve the limits to cosmic opacity at high redshifts ($z>2$) as well as test new parameterizations for $\tau(z)$, which can help us to
pinpoint the nature of dark energy.

\label{sec:conclusions}

\section*{Acknowledgments}

R.F.L.H thanks INCT-A and is supported by CNPq (No. 478524/2013-7). V.C.B is supported by CNPq-Brazil through a fellowship inside the program Science without Borders.


\begin{thebibliography}{99}

\bibitem{caldwell} R. R. Caldwell and M. Kamionkowski, Ann.\ Rev.\ Nucl.\ Part.\ Sci. {\bf 59}, 525 (2009).

\bibitem{li} M. Li, X.-D. Li, S. Wang, and Y. Wang, Commun. Theor. Phys. {\bf 56}, 397 (2011).

\bibitem{Lima1999} J. A. S. Lima and J. S. Alcaniz, Astron. Astrophys. {\bf 348}, 1 (1999).

\bibitem{chimento} L. P. Chimento, A. S. Jakubi, and D. Pav\'{o}n, Phys. Rev. D {\bf 67}, 087302 (2003).

\bibitem{drell} P. S. Drell, T. J. Loredo, and I. Wasserman, Astrophys. J. {\bf 530}, 593 (2000).

\bibitem{combes} F. Combes, New Astronomy Rev. {\bf 48}, 583 (2004).

\bibitem{zehavi} I. Zehavi {\it et al.}, Astrophys. J. {\bf 503}, 483 (1998).

\bibitem{conley} A. Conley {\it et al.}, Astrophys. J. {\bf 664}, L13 (2007).

\bibitem{Ishak} M. Ishak, A. Upadhye, and D. N. Spergel, Phys. Rev. D {\bf 74}, 043513 (2006).

\bibitem{ks2007} M. Kunz and D. Sapone, Phys. Rev. Lett. {\bf 98}, 121301 (2007).

\bibitem{Bertschinger} E. Bertschinger and P. Zukin, Phys. Rev. D {\bf 78} 024015 (2008).

\bibitem{Aguirre} A. Aguirre, Astrophys. J.{\bf 525}, 583 (1999).

\bibitem{Rowan} M. Rowan-Robinson, Mon. Not. R. Astron. Soc. {\bf 332}, 352 (2002).

\bibitem{Goobar} A. Goobar, L. Bergstrom, and E. Mortsell, Astron. Astrophys. {\bf 384}, 1 (2002).

\bibitem{Av2} A. Avgoustidis, C. Burrage, J. Redondo, L. Verde, and
R. Jimenez, J. Cosmol. Astropart. Phys. {\bf 10},  024 (2010). 

\bibitem{Jaeckel} J. Jaeckel and A. Ringwald, Ann. Rev. Nucl. Part.
Sci. {\bf 60}, 405 (2010).

\bibitem{eth33} I. M. H. Etherington,  Philos. Mag. {\bf 15}, 761 (1933).

\bibitem{ellis07} G. F. R. Ellis, Gen. Relativ. Gravit {\bf 39}, 1047 (2007).

\bibitem{busti2012} V. C. Busti and J. A. S. Lima,  Mon. Not. R. Astron. Soc. {\bf 426}, L41 (2012).

\bibitem{clarkson} C. Clarkson, G. F. R. Ellis, A. Faltenbacher, R. Maartens, O. Umeh, and J.-P. Uzan, Mon. Not. R. Astron. Soc. {\bf 426}, 1121 (2012).

\bibitem{bk04} B. A. Bassett and M. Kunz, Phys. Rev. D {\bf 69}, 101305 (2004).

\bibitem{daly} R. A. Daly and S. G. Djorgovski, Astrophys. J. {\bf 597}, 9 (2003).

\bibitem{G94} L. I. Gurvits, Astrophys. J. {\bf 425}, 442 (1994).

\bibitem{h3} R. F. L. Holanda, J. A. S. Lima, and M. B. Ribeiro, Astrophys. J. {\bf 722}, L233 (2010).

\bibitem{h2} R. F. L. Holanda, J. A. S. Lima, and M. B. Ribeiro, Astron. Astrophys. {\bf 528}, L14 (2011).

\bibitem{h3a} R. F. L. Holanda, J. A. S. Lima, and M. B. Ribeiro, Astron. Astrophys. {\bf 538}, 131 (2012).

\bibitem{Nair} R. Nair, S. Jhingan, and D. Jain, J. Cosmol. Astropart. Phys. {\bf 05} (2011) 023.

\bibitem{xi} Y. Xi, Y. Hao-Ran, Z. Zhi-Song, and Z. Tong-Jie, Astrophys. J. {\bf 777}, L24 (2013).

\bibitem{nam} L. Nam {\it et al.}, Mon. Not. R. Astron. Soc. {\bf 436}, 1017 (2013).

\bibitem{li2013} Z. Li, P. Wu, H. Yu, and Z.-H. Zhu, Phys. Rev. D {\bf 87}, 103013 (2013). 

\bibitem{Amanullah} R. Amanullah {\it et al.}, Astrophys. J. {\bf 716}, 712 (2010).

\bibitem{Lima} J. A. S. Lima, J. V. Cunha, and V. T. Zanchin, Astrophys. J. {\bf 742}, L26 (2012).

\bibitem{HAC} { R. F. L. Holanda, J. C. Carvalho, and J. S. Alcaniz, J. Cosmol. Astropart. Phys.  {\bf 04}, (2013) 027.} 

\bibitem{foot} { The opacity-free luminosity distances  were obtained from a numerical integration of current $H(z)$ data points and not in the context of a given cosmological model.}

\bibitem{ellis2013} G. F. R. Ellis, R. Poltis, J.-P. Uzan, and A. Weltman, Phys. Rev. D {\bf 87}, 103530 (2013).

\bibitem{h5b} R. F. L. Holanda, R. S. Gon\c{c}alves, and J. S. Alcaniz, J. Cosmol. Astropart. Phys. {\bf 06} (2012) 022.

\bibitem{amati} L. Amati {\it et al.}, Astron. Astrophys. {\bf 390}, 81 (2002).

\bibitem{ghirlanda} G. Ghirlanda, G. Ghisellini, D. Lazzati, and C. Firmani, Astrophys. J. {\bf 613}, L13 (2004).

\bibitem{schaefer2007} B. E. Schaefer, Astrophys. J. {\bf 660}, 16 (2007).

\bibitem{wang} S. Wang and Y. Zhang, Phys. Lett. B {\bf 669}, 201 (2008).

\bibitem{ratra2010} L. Samushia and B. Ratra, Astrophys. J. {\bf 714}, 1347 (2010).

\bibitem{wei2} J. J. Wei, X. Wu, and F. Melia, Astrophys. J. {\bf 772}, 43 (2013).

\bibitem{busti_prd_2012} V. C. Busti, R. C. Santos, and J. A. S. Lima, Phys. Rev. D {\bf 85}, 103503 (2012). 

\bibitem{Av} A. Avgoustidis, C. Burrage, J. Redondo, L. Verde, and
R. Jimenez, J. Cosmol. Astropart. Phys. 06 (2009) 012.

\bibitem{chen2009a} B. Chen and R. Kantowski, Phys. Rev. D {\bf 79}, 104007 (2009).

\bibitem{chen2009b} B. Chen and R. Kantowski, Phys. Rev. D {\bf 80} 044019 (2009).

\bibitem{Hogg} D. W. Hogg, arXiv:astro-ph/9905116 (1999).

\bibitem{footnote} Naturally, in our case, the equivalence between CDD relation and cosmic opacity cannot be used since the GRB data lie in $z>2$. Our choice for a linear parameterization is due to its simplicity and the low quality of the available GRB data, which does not allow constraints for more general expressions.

\bibitem{suzuki} N. Suzuki {\it et al.}, Astrophys. J. {\bf 746}, 24 (2012). 

\bibitem{Rubin} D. Rubin {\it et al.}, Astrophys. J. {\bf 763}, 35 (2013).

\bibitem{highzgrb} { R. Salvaterra {\it et al.}, Nature (London) {\bf 461}, 1258 (2009).}

\bibitem{highzgrb2} A. Cucchiara {\it et al.}, Astrophys. J. {\bf 736}, 7 (2011).

\bibitem{kodama} Y. Kodama, D. Yonetoku, T. Murakami, S. Tanabe, R.
Tsuitsui, and T. Nakamura, Mon. Not. R. Astron. Soc. {\bf 391}, L1 (2008).

\bibitem{liang} N. Liang, W. K. Xiao, Y. Liu, and S. N. Zhang, Astrophys.
J. 685, 354 (2008).

\bibitem{wei} H. Wei, J. Cosmol. Astropart. Phys. {\bf 08} (2010) 020.

\bibitem{atteia} V. Heussaff, J.-L. Atteia, and Y. Zolnierowski, Astron. Astrophys. {\bf 557}, A100 (2013).

\bibitem{salvaterra} { R. Salvaterra {\it et al.}, Astrophys. J. {\bf 749}, 68 (2012).}

\bibitem{simon} J. Simon, L. Verde, and R. Jimenez, Phys. Rev. D {\bf 71}, 123001 (2005).

\bibitem{stern} D. Stern {\it et al.}, Astrophys. J. Supp. Ser. {\bf 188}, 280 (2010).

\bibitem{chen2011} G. Chen and B. Ratra, PASP {\bf 123}, 1127 (2011).

\bibitem{seik} M. Seikel, S. Yahya, R. Maartens, and C. Clarkson, Phys. Rev. D {\bf 86}, 083001 (2012).

\bibitem{moresco} M. Moresco {\it et al.}, J. Cosmol. Astropart. Phys. {\bf 08} (2012) 006.

\bibitem{Farooq} O. Farooq, D. Mania, and B. Ratra, Astrophys. J. {\bf 764}, 138 (2013).

\bibitem{blake} C. Blake {\it et al.}, Mon. Not. R. Astron. Soc. {\bf 425}, 405 (2012).

\bibitem{reid} B. A. Reid {\it et al.}, Mon. Not. R. Astron. Soc. {\bf 426}, 2719 (2012).

\bibitem{xu} X. Xu, A. J. Cuesta, N. Padmanabhan, D. J. Eisenstein, and C. K. McBride, Mon. Not. R. Astron. Soc. {\bf 431}, 2834 (2013).

\bibitem{busca} N. G. Busca  {\it et al.}, Astron. Astrophys. {\bf 552}, 96 (2013).

\bibitem{chuang} C.-H. Chuang, and Y. Wang, Mon. Not. R. Astron. Soc. {\bf 435}, 255 (2013).

\bibitem{Planck} Planck Collaboration, arXiv:1303.5076 (2013).

\bibitem{guy} J. Guy {\it et al.}, Astron. Astrophys. {\bf 466}, 11 (2007).

\bibitem{riess} A. G. Riess {\it et al.}, Astrophys. J. {\bf 607}, 665 (2004).

\bibitem{h5} R. S. Gon\c{c}alves, R. F. L. Holanda, and J. S. Alcaniz, Mon. Not. R. Astron. Soc. {\bf 420}, L43 (2012).

\bibitem{jimenez} R. Jimenez and A. Loeb, Astrophys. J. {\bf 573}, 37 (2002).	

\bibitem{Meng} X. Meng, T. Zhang, H. Zhan, and X. Wang, Astrophys. J. {\bf 745}, 98 (2012).











\end{thebibliography}
\end{document}